\documentclass[12pt]{article}

\input epsf.sty

\newcommand{\insertfig}[2]{\mbox{\epsfxsize=#1cm \epsfbox{#2.eps}}}
\newcommand{\ft}[2]{{\textstyle\frac{#1}{#2}}}

\font\cmss=cmss12 
\def\1{\hbox{{1}\kern-.25em\hbox{l}}}
\def\bfZ{\relax{\hbox{\cmss Z\kern-.4em Z}}}

\begin{document}
\begin{titlepage}

\begin{flushright}
YITP-SB-01-25\\ [-2mm]
LPT-Orsay-01-60 \\ [-2mm]
BNL-HET-01/24
\end{flushright}

\centerline{\large \bf Energy flow in QCD and event shape functions. }

\vspace{15mm}

\centerline{\bf A.V. Belitsky$^{a,b}$, G.P. Korchemsky$^b$, G. Sterman$^{a,c}$}

\vspace{15mm}

\centerline{\it $^a$C.N.\ Yang Institute for Theoretical Physics}
\centerline{\it State University of New York at Stony Brook}
\centerline{\it NY 11794-3840, Stony Brook, USA}

\vspace{5mm}

\centerline{\it $^b$Laboratoire de Physique Th\'eorique\footnote{Unite Mixte 
                de Recherche du CNRS (UMR 8627).},
                Universit\'e de Paris XI}
\centerline{\it 91405 Orsay C\'edex, France}

\vspace{5mm}

\centerline{\it $^c$Physics Department, Brookhaven National Laboratory}
\centerline{\it NY 11973-5000, Upton, USA}

\vspace{15mm}

\centerline{\bf Abstract}

\vspace{0.5cm}

\noindent

Hadronization corrections to the thrust and related event shape distributions 
in the two-jet kinematical region of ${\rm e}^+{\rm e}^-$ annihilation are 
summarized by nonperturbative shape functions. The moments of shape functions 
are given by universal matrix elements in QCD, which describe the energy flow 
in QCD final states.  We show how the nonperturbative structure of these 
matrix elements may be inferred from resummed perturbation theory and Lorentz 
invariance. This analysis suggests the same functional forms for the shape 
functions as were found in phenomenological studies, and sheds light on the 
physical significance of the parameters that characterize these functions.

\vspace{1.5cm}



\end{titlepage}

{\bf 1. Introduction.} The predictive power of perturbative QCD is based on
the factorization of long- and short-distance dynamics, made possible
by their quantum-mechanical incoherence. The hard scattering of quarks and 
gluons can be computed in ordinary perturbation theory, while soft-scale 
physics  is encoded into nonperturbative functions, which are fit to data, 
or evaluated using models. In a rather limited number of inclusive processes, 
which admit the operator product expansion (OPE), these functions are 
described by universal inclusive distributions, for example, leading-twist 
parton densities in deeply inelastic scattering.  The latter can be defined 
in QCD in terms of correlators of quark and gluon fields on the light cone, 
and can, in principle, be computed using nonperturbative methods. On the 
other hand, there is a large variety of processes that are not completely 
inclusive, but are nevertheless infrared safe \cite{SteWei78}, and thus 
calculable in perturbation theory. Important examples are event shape 
distributions in ${\rm e}^+ {\rm e}^-$ final states \cite{Bie01}. 
Nonperturbative physics enters these cross sections as corrections, 
suppressed by powers of the  large energy scale, $Q$.

Since event shape distributions are weighted, rather than fully inclusive, 
cross sections, their power corrections are not constrained by the OPE, and 
are generally less suppressed.  For instance, the mean value of the thrust, 
$T$, acquires a nonperturbative correction $\langle t \rangle_{\rm NP} \sim 
{\mit\Lambda}/Q$ with $t = 1 - T$, and ${\mit\Lambda}$ a new QCD scale 
\cite{IRR}. We may think of this correction as the first term in an expansion 
in $1/Q$. The situation is even more complex for differential event shape 
distributions, such as $d\sigma/dt$, in the two-jet kinematic region, 
$t\rightarrow 0$.  In this case, one has to deal with an infinite number of 
scales, $\left( d \sigma/ dt \right)_{\rm NP} \sim \sum_{k} {\mit\Lambda_k}/
\left(t Q\right)^k$. The organization of such a series is equivalent to the 
introduction of a new nonperturbative distribution.  These distributions are 
the shape functions \cite{Kor98,KorSte99}. They provide a successful
phenomenological description of differential event shapes, down to the 
two-jet limit, and over a wide interval of energies, as shown in Refs.\ 
\cite{KorTaf00,GarRat01}.  In this paper, we show how the functional form 
found in these phenomenological studies may be motivated from QCD.

A perturbative analysis of the event shape distributions in the two-jet
region shows that the leading power corrections, $1/(tQ)^k$, are associated
with multiple soft, wide-angle gluon emission at the momentum scale $t Q$ 
\cite{Kor98,KorSte99}. The influence of collinear splittings of quarks and 
gluons is suppressed by an extra power of $Q$. The leading behavior of the 
cross section is thus independent of the dynamics that produces the
internal structure of jets, and is well described by a physical picture of 
the perturbative final state in which two fast moving back-to-back quarks 
propagate through a cloud of soft gluons, behaving as classical sources of
color charge.  This is equivalent to the eikonal approximation
for the quark-antiquark pair.

In eikonal approximation, multiple soft-gluon emission from quarks is 
described by nonabelian phase operators ${\mit\Phi}_{n_R} 
[ \infty, y ]$ and ${\mit\Phi}^\dagger_{n_L} [ \infty, y ]$. Here
${\mit\Phi}_{n} [ \infty, y ] = P \exp \left( i g
\int_{0}^{\infty} d \sigma \, n^\mu A_\mu (\sigma n + y) \right)$
is a Wilson line, with its light-like direction $n_\mu$ defined by a 
quark momentum, $p_1^\mu =  n_L^\mu \sqrt{Q^2/2}$ or $p_2^\mu = n_R^\mu 
\sqrt{Q^2/2}$, with $Q^2 = (p_1 + p_2)^2$ the total center-of-mass energy. 
In a frame where the two-jet axis points in the $z$-direction, we choose 
the four-velocities $n^\mu_R = \ft1{\sqrt{2}} (1, \mbox{\boldmath$0$}, 1)$ 
and $n^\mu_L = \ft1{\sqrt{2}} (1, \mbox{\boldmath$0$}, - 1)$, with $n_L^2 
= n_R^2 = 0$ and $n_L \cdot n_R = 1$. Combining the eikonal phases of 
quark and antiquark, we obtain
\begin{eqnarray}
U (y ; A) =
T \left\{ {\mit\Phi}_{n_R} [ \infty, y ]
{\mit\Phi}^\dagger_{n_L} [ \infty , y ] \right\}
= T \left\{ {\mit\Phi}_{n_R} [ \infty, y ]
{\mit\Phi}_{n_L} [ y, \infty ] \right\}\, ,
\end{eqnarray}
where we have used the unitarity of the phase operators. The symbol $T$ 
stands for time ordering of the gauge fields $A_\mu$ in the product. The 
explicit expression for $U$ can then be written as
\begin{eqnarray}
\label{Pexponent}
U (y ; A)
= T P \exp
\left(
i g \int_{C_{{\rm e}^+ {\rm e}^-}} d z^\mu A_\mu (y + z)
\right)\, ,
\end{eqnarray}
where $P$ orders the color indices of gluon fields along the integration path
$C_{{\rm e}^+ {\rm e}^-}$: $z^\mu (\tau) = \tau \, n^\mu_R \, \theta (- \tau)
+ \tau \, n^\mu_L \, \theta (\tau)$ with $- \infty < \tau < \infty$.

A characteristic of weighted cross sections like the thrust and the
heavy jet mass is that they distinguish between gluons moving into the right 
and left hemispheres, defined with respect to the plane orthogonal to the 
two-jet axis (in the overall center-of-mass frame). When the quarks are 
treated
in the eikonal approximation, the shape function that describes energy flow 
for
these final states is \cite{Kor98,KorSte99}
\begin{equation}
\label{NonOPEf}
f ( \varepsilon_R, \varepsilon_L )
= \sum_{N} |\langle 0 | U (0) | k_1, \dots , k_N \rangle |^2
\delta \left( \varepsilon_R - \sum_{i \in R} k_{i,+} \right)
\delta \left( \varepsilon_L - \sum_{j \in L} k_{j,-} \right)\, .
\end{equation}
Here, the summation goes over soft particles in the final states, with
momenta $k_1, \dots , k_N$. The variables in the shape function,
$\varepsilon_R$ and $\varepsilon_L$, are projections of the total momenta of
soft particles flowing into the right and left hemispheres onto the 
corresponding jet directions,
\begin{equation}
k_+ \equiv k \cdot n_R, \qquad\qquad k_- \equiv k \cdot n_L\, .
\end{equation}
The operator $U(0)$ in Eq.\ (\ref{NonOPEf}) represents
radiation from the quark-antiquark pair, in the eikonal
approximation discussed above.

\begin{figure}[t]
\begin{center}
\hspace{0cm} \mbox{
\begin{picture}(0,125)(100,0)
\put(10,-14){\insertfig{6}{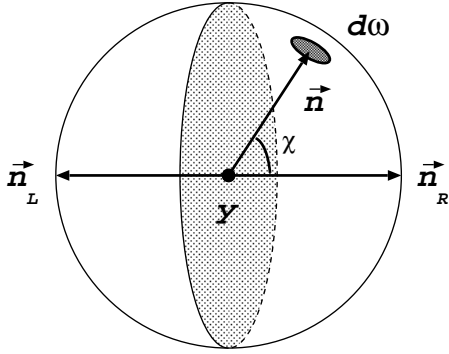}}
\end{picture}
}
\end{center}
\caption{\label{Sphere} The pictorial definition of the energy flow operator
${\cal E} (\vec n)$. The unit vectors $\vec n_R$ and $\vec n_L$
indicate the directions of the two outgoing jets. The shaded plane orthogonal
to these unit vectors goes through the annihilation point, and separates the
left and right hemispheres. The energy flux through the infinitesimal surface
element $d \omega = d\vec n \, \delta \left( \vec n^2 - 1 \right)$ is given by
${\cal E} (\vec n) d \omega$.}
\end{figure}

To derive a practical operator interpretation for the event shape function
(\ref{NonOPEf}) in QCD, it is convenient to eliminate the summation over final
states. To this end, we introduce an energy flow operator, ${\cal E}
(\vec n)$, which acts on asymptotic states by measuring
the differential distribution of particles with energy $k^0$
moving in the solid angle $d \vec n \, \delta (\vec n^2 - 1) = d \cos
\chi d \varphi$ \cite{OreSte80,SveTka96,KorOdeSte97}.
We define this operator by its action on a final
state consisting of $N$ particles with momenta $k_1, \dots, k_N$,
according to
\begin{equation}
\label{ElguonState}
{\cal E} (\vec n) | k_1, \dots , k_N \rangle
= \sum_{j = 1}^{N} k^0_j \,
\delta ( \cos \theta_j - \cos \chi ) \delta (\phi_j - \varphi)
| k_1, \dots , k_N \rangle\,  ,
\end{equation}
where spherical angles $(\theta_j, \phi_j)$ define the position of
the unit vector $\vec n_j = \vec k_j/|\vec k_j|$ with respect to the jet
axis $\vec n_R$, and $(\chi, \varphi)$ define those of $\vec n$.
Introducing a light-like vector $n^{\mu} = (1, \vec n)$, we can
represent the total momentum flow through the right ($n^3 > 0$) and
left ($n^3 < 0$) hemispheres of infinite radius in terms of
\begin{eqnarray}
{\cal P}^\mu_{R,L} = \int d \vec n \, n^\mu \, \theta (\pm n^3)
\delta \left( \vec n^2 - 1 \right) {\cal E} (\vec n)\, ,
\end{eqnarray}
so that
\begin{equation}
{\cal P}^\mu_R | k_1, \dots , k_N \rangle
= \sum_{j \in R} k^\mu_j | k_1, \dots , k_N \rangle ,
\end{equation}
and similarly for ${\cal P}^\mu_L$. Notice that the total momentum of the
final state is given by the sum ${\cal P}^\mu = \int d \vec x \
{\mit\Theta}^{0\mu} (t, \vec x) = {\cal P}^\mu_R + {\cal P}^\mu_L$,
with ${\mit\Theta}^{\mu\nu}$ the energy-momentum tensor.

Using these definitions, the shape function, Eq.\ (\ref{NonOPEf}),
can be reexpressed as
\begin{equation}
\label{ShapeFuncTwo}
f (\varepsilon_R, \varepsilon_L)
= \langle 0 |
U^\dagger (0)
\delta
\left( \varepsilon_R - {\cal E}_R \right)
\delta
\left( \varepsilon_L - {\cal E}_L \right)
U (0)
| 0\rangle\, ,
\end{equation}
where
\begin{equation}
\label{ERL}
{\cal E}_{R,L} = \int d \vec n \, \delta \left( \vec n^2 - 1 \right)
w_{R,L} (\vec n) {\cal E} (\vec n)\, ,
\qquad\qquad
w_{R,L} (\vec n) = \left( 1 - |\cos \chi| \right)
\theta \left( \pm \cos \chi \right)\, .
\end{equation}
Defined in this way, the shape function becomes a new QCD distribution, 
which governs nonperturbative power corrections to a number of differential 
cross sections in the two-jet region.

A typical application of the shape function, which
illustrates the factorization of soft gluon
emission from perturbative dynamics, is to the
distribution for the heavy jet mass.
The heavy jet mass is defined by $\varrho= (1/Q^2){\rm max}(m_R^2,m_L^2)$,
with $m_R$ ($m_L$) the invariant mass of the particles moving into
the right (left) hemisphere.
It may be written in the convolution form \cite{KorSte99},
\begin{equation}
\label{convolution}
\frac{1}{\sigma_{\rm tot}} \frac{d \sigma_\varrho}{d \varrho}
= Q f_\varrho \left( \varrho Q, \varrho Q \right)
{\cal R}^{\rm PT}_J (0)
+ \int_{0}^{\varrho Q} d \varepsilon
f_\varrho \left( \varepsilon, \varrho Q \right)
\frac{d \sigma^{\rm PT}_J
\left( \varrho - \ft{\varepsilon}{Q} \right)}{d \varrho}\, ,
\end{equation}
where $d \sigma^{\rm PT}_J / d \varrho$ is the resummed perturbative
single-jet cross section \cite{CTTW}, ${\cal R}^{\rm PT}_J$ is the 
corresponding `radiator function', defined by $d {\cal R}^{\rm PT}_J 
(\varrho) / d \varrho = d \sigma^{\rm PT}_J / d \varrho$, and
\begin{equation}
f_\varrho (\varepsilon, \varrho Q) = 2 \int_{0}^{\varrho Q}
d \varepsilon' f \left( \varepsilon, \varepsilon' \right)
{\cal R}^{\rm PT}_J \left( \varrho - \frac{\varepsilon'}{Q} \right) \, .
\end{equation}
The thrust and the $C$-parameter distributions are given by similar
expressions \cite{KorTaf00}, where $f_\varrho$ is replaced by a 
single-variable function, related to $f (\varepsilon_R, \varepsilon_L)$ 
by\footnote{In what follows we use a convention $\langle 0 
| U^\dagger \dots U | 0 \rangle \equiv \langle \dots \rangle$.}
\begin{equation}
f_t (\varepsilon) = \int_{0}^{\infty}
d \varepsilon_R d \varepsilon_L \ f (\varepsilon_R, \varepsilon_L)
\, \delta \left( \varepsilon - \varepsilon_L - \varepsilon_R \right)
= \langle
\delta \left( \varepsilon - {\cal E}_R - {\cal E}_L \right)
\rangle \, .
\label{ft}
\end{equation}
Corrections to Eq.\ (\ref{convolution}) are suppressed by powers
of $Q$, as discussed above.

In Ref.\ \cite{KorTaf00}, it was found that a successful description
of the differential distributions for thrust, heavy jet mass, and 
$C$-parameter
over a wide range of energies is given by the single shape function,
\begin{equation}
f(\varepsilon_R,\varepsilon_L)
= {{\cal N}(a,b)\over \Lambda^2}\,
\left( {\varepsilon_R\varepsilon_L\over \Lambda^2} \right)^{a-1}\;
\exp\left( - {\varepsilon_R^2 +\varepsilon_L^2 + 2b\varepsilon_R\varepsilon_L
\over \Lambda^2} \right)\, ,
\label{KTansatz}
\end{equation}
when its parameters are chosen as: $a=2,\ b=-0.4$ and $\Lambda=0.55$ GeV.
The normalization factor ${\cal N}(a,b)$ is chosen so that
$\int d\varepsilon_R d\varepsilon_L f(\varepsilon_R,\varepsilon_L)=1$.
Recently, a thorough investigation of the differential distribution
for thrust \cite{GarRat01} reached similar conclusions in terms of
the function $f_t$, Eq.\ (\ref{ft}). In the following, we shall argue that 
the functional form, Eq.\ (\ref{KTansatz}), follows from a perturbative 
analysis of moments of the matrix element (\ref{ShapeFuncTwo}), supplemented 
by Lorentz invariance and standard treatment of integrals of the strong 
coupling over low momentum scales.

{\bf 2. Energy correlators.} In a manner analogous to the relation 
between the light-cone expansion and moments of deep-inelastic scattering 
structure functions, the moments of event shape functions
give matrix elements of certain (in this case, nonlocal) operators in
QCD. Specifically, from the definition, Eq.\ (\ref{ShapeFuncTwo}),
the moments of $f (\varepsilon_R, \varepsilon_L)$ produce weighted Green
functions of products of the operators ${\cal E}$, inserted
into correlators of Wilson lines,
\begin{eqnarray}
\label{MomentsEREL}
&&\int_0^\infty d \varepsilon_R d \varepsilon_L \
\varepsilon_R^N \varepsilon_L^M \ f (\varepsilon_R, \varepsilon_L)
= \langle {\cal E}_R^N {\cal E}_L^M \rangle
\\
&&\qquad\qquad\qquad\qquad=
\int \prod_{j = 1}^{N + M} d \vec n_j \,
\delta \left( \vec n_j^2 - 1 \right)
\prod_{n = 1}^{N} w_R (\vec n_n)
\prod_{m = N + 1}^{M} w_L (\vec n_m)
{\cal G} (\vec n_1 , \dots , \vec n_{N + M})\, ,
\nonumber
\end{eqnarray}
where
\begin{eqnarray}
\label{GreenFunction}
{\cal G} (\vec n_1, \dots , \vec n_N)
\!\!\!&=&\!\!\!
\langle {\cal E} (\vec n_1) \dots {\cal E} (\vec n_N) \rangle \\
&\equiv&\!\!\!
\frac{1}{N_c}
\langle 0 |
{\rm Tr}
\left\{ U^\dagger \left( 0; A^{(-)} \right)
{\cal E} (\vec n_1) \dots {\cal E} (\vec n_N)
U \left( 0; A^{(+)} \right) \right\}
| 0 \rangle\, \nonumber
\end{eqnarray}
measures correlations between energy flows in the directions of
specified unit vectors $\vec n_j$. Rather than study the shape function
directly at first, we begin with these, even more general matrix elements,
from which the shape function (\ref{ShapeFuncTwo}) may be derived
\cite{KorSte99}.

We shall analyze below the energy correlators, ${\cal G} (\vec n_1, \dots , 
\vec n_N)$, from a general point of view that has been widely applied to 
infrared safe quantities, using a variety of related prescriptions 
\cite{IRR}. This may be summarized very briefly as follows. Starting 
with low-order diagrams for the quantity in question, we absorb 
higher-order corrections into the scale of the running coupling. In the 
resulting expression, the integral over the argument of the running
coupling from zero up to a factorization scale, $\mit\Lambda$,
is replaced by a phenomenological parameter.  Other, overall kinematic 
integrals, upon which the running coupling does not depend, determine
properties like angular and rapidity  dependence. We will show that 
this procedure, applied at order $\alpha_s$ and $\alpha_s^2$
to the energy correlators, is adequate to derive the functional form Eq.\ 
(\ref{KTansatz}) for the corresponding shape functions, and  to give
physical interpretations to its dimensionless parameters $a$ and $b$.

\begin{figure}[t]
\begin{center}
\hspace{0cm}
\mbox{
\begin{picture}(0,125)(100,0)
\put(10,-14){\insertfig{6}{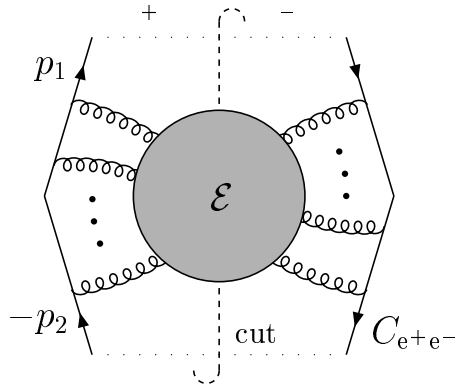}}
\end{picture}
}
\end{center}
\caption{\label{GreenFigure} Diagrammatic representation of the
correlation function (\ref{GreenFunction}). Here the dashed line
represents the unitary cut and the final states are weighted with
the product of factors ${\cal E} (\vec n_1) \dots {\cal E} (\vec n_N)$
defined in Eq.\ (\ref{ElguonState}).}
\end{figure}

The perturbative expansions of the correlators of Eq.\ (\ref{GreenFunction})
are illustrated diagrammatically by Fig.\ \ref{GreenFigure}. As the figure 
shows, these correlation functions correspond to cut diagrams, rather than 
time-ordered Green functions, since the gauge fields entering the eikonal 
phases $U$ and $U^\dagger$ are time and anti-time ordered, respectively. 
To express this fact, we adopt the Keldysh notation, assigning ``plus" 
and ``minus" superscripts to the fields entering the amplitude ($A^{(+)}$) 
and its conjugated counterpart ($A^{(-)}$). The propagator of the gauge fields
$A^{(\pm)}$ is defined  as
\begin{eqnarray}
\int d^4 x \, {\rm e}^{i q \cdot x}
\langle 0 |
A_\mu^{(\alpha)} (x) A_\nu^{(\beta)} (0)
| 0 \rangle
= - i \varrho_{\mu\nu} (q) D^{(\alpha \beta)} (q) ,
\end{eqnarray}
with $\alpha, \beta = \pm$, and with the conventional
free gluon polarization tensor in covariant gauge, $\varrho_{\mu\nu}
(q) = g_{\mu\nu} - (1 - \xi) q_\mu q_\nu / q^2$.
In these terms, we have $D^{(++)} (q) = - [D^{(--)} (q)]^\star = (q^2 +
i 0)^{-1}$, $D^{(-+)} (q) = D^{(+-)} (- q) = - 2 \pi i \theta (q^0) \delta
(q^2)$. We choose below the Feynman gauge, $\xi = 1$, for calculations
(although our results are gauge invariant).

{\bf 3. The single-gluon approximation}.
We begin our study of the properties of the correlation functions
with the lowest-order perturbative expression for the single
energy correlator, ${\cal G} (\vec n)$. Using the definition
(\ref{GreenFunction}) and expanding (\ref{Pexponent}) in powers
of the gauge coupling, we have
\begin{equation}
{\cal G} ( \vec n )
= g^2 \int_{C_{{\rm e}^+ {\rm e}^-}} \! dz^\mu_1 \int_{C_{{\rm e}^+ {\rm
e}^-}}
\! dz^\nu_2 \ \langle 0 | \,
A^{(-)}_\mu (z_1) {\cal E} (\vec n) A^{(+)}_\nu (z_2) \,
| 0 \rangle
+ {\cal O} \left( g^4 \right) .
\end{equation}
Going to the momentum representation, and performing the integration
along the contour, we find
\begin{eqnarray}
\label{1G}
{\cal G} ( \vec n )
= \frac{C_F \alpha_s}{2 \pi^2}
\int \frac{dk_+}{k_+} \int \frac{d \mbox{\boldmath$k$}{}^2}{
\mbox{\boldmath$k$}{}^2} \, \int d\phi_k \,
\, k_0 \, \delta ( \cos \theta_k - \cos \chi ) \delta (\phi_k - \varphi)\, .
\end{eqnarray}
In this formula, we recognize the lowest-order bremsstrahlung
spectrum of soft gluons, weighted by an additional energy flow factor,
according to Eq.\ (\ref{ElguonState}). Using the definitions of the light-cone
variables $k_\pm = \ft1{\sqrt{2}} \left( k_0 \pm k_3 \right)$,
along with $k_0 = \sqrt{\mbox{\boldmath$k$}{}^2}
\sin^{-1}\! \theta_k$, $k_3 = \sqrt{\mbox{\boldmath$k$}{}^2} \cot
\theta_k$, we find $dk_+/k_+ = d \cos \theta_k / \sin^2 \theta_k$.
The integration over $k_+$ then yields the result
\begin{eqnarray}
\label{1G-int}
{\cal G} ( \vec n )
= \frac{1}{2 \pi} \frac{1}{\sin^3\! \chi}
\int d \mbox{\boldmath$k$}{}^2 \, \sqrt{\mbox{\boldmath$k$}{}^2}
\, \rho_{\rm PT} \left( \mbox{\boldmath$k$}{}^2 \right)\, ,
\label{Gn}
\end{eqnarray}
where we have introduced a resummed perturbative density, $\rho_{\rm PT} 
\left( \mbox{\boldmath$k$}{}^2 \right) \equiv \frac{C_F}{\pi}
\frac{\alpha_s (\mbox{\boldmath${\scriptstyle k}$}{}^2)}{\mbox{\boldmath
${\scriptstyle k}$}{}^2}$ \cite{IRR}, which measures the number of particles 
produced per unit interval in rapidity and transverse momentum.
The expression (\ref{1G-int}) exhibits a third-order pole as $\chi \to 0$ or
$\chi \to \pi$. In both cases, a gluon propagates in the final state close to
the direction of an outgoing jet, and the resulting singularities are of 
collinear origin.

The simplicity of the  $\chi$-dependence in Eq.\ (\ref{Gn}) reflects the
underlying boost-invariance of the nonabelian phase operators $U$.
All dynamical information is contained in
a boost-invariant particle density, independent of the rapidity,
\begin{equation}
\label{Rapidity}
\eta \equiv \frac{1}{2} \ln \frac{k_+}{k_-} = - \ln \tan \frac{\chi}{2}\, .
\label{etachi}
\end{equation}
The resummation incorporated into the
running coupling in the step from Eq.\ (\ref{1G}) to Eq.\ (\ref{1G-int})
is the only choice consistent with the boost properties of the correlators.
This resummation, which summarizes higher-order scale-fixing corrections, 
$\alpha_s\left(\beta_0\, \alpha_s \right)^N$, is normally justified 
diagrammatically, and is used as a starting point for infrared 
renormalon-inspired arguments \cite{IRR}.  We shall not employ the
form of the perturbative running coupling, however, nor will we make a 
specific ansatz for a nonperturbative generalization of the integrals in
Eq.\ (\ref{1G-int}). Following Refs.\ \cite{IRR}, however, we replace the
resummed perturbative density by a phenomenological one, $\rho_{\rm NP}
(\mbox{\boldmath$k$}{}^2)$, below some cutoff, ${\mit\Lambda}^2$, in the 
$\mbox{\boldmath$k$}{}^2$ integral. Whatever the choice of $\rho_{\rm NP}$, 
it should, like its perturbative counterpart, be boost invariant.
For any such choice of the nonperturbative density, the single-particle 
energy correlator, ${\cal G} (\vec n)$, defines a leading-power correction 
to the energy-energy correlation in ${\rm e}^+ {\rm e}^-$-annihilation
\cite{BasBroEllLov79},
\begin{equation}
\langle \mbox{EEC} (\chi) \rangle_{\rm NP} = \frac{1}{Q} {\cal G} (\vec n)\, ,
\end{equation}
with the rapidity dependence specified by Eqs.\ (\ref{Gn}) and (\ref{etachi}),
and any choice also produces a uniform particle number
distribution, $d^2 N / d \eta d \mbox{\boldmath$k$}{}^2 = \rho 
(\mbox{\boldmath$k$}{}^2)$, in terms
of the rapidity.

We have thus, starting with the energy correlators, arrived at a ``tube" 
model, as reviewed in Ref.\ \cite{Web94}, in which  particles are produced with
a constant, nonperturbative density at all rapidities. The perturbative 
calculation therefore suggests a realistic dependence on the rapidity,
in terms of an integral of the density $\rho$ over nonperturbative scales.  
This integral, of course, is to be replaced by a set of phenomenological
parameters. We now use this approach in the computation of multiple energy
correlators.  Perhaps surprisingly, the lowest-order model offers
insights here as well.

To the lowest order in perturbation theory, the multiple energy
flow correlator ${\cal G} (\vec n_1, \dots , \vec n_j)$ is given by
\begin{equation}
{\cal G} ( \vec n_1, \dots , \vec n_j )
= g^2 \int_{C_{{\rm e}^+ {\rm e}^-}} \! dz^\mu_1
\int_{C_{{\rm e}^+ {\rm e}^-}} \! dz^\nu_2
\ \langle 0 | \,
A^{(-)}_\mu (z_1)
{\cal E} (\vec n_1) \dots  {\cal E} (\vec n_j)
A^{(+)}_\nu (z_2) \,
| 0 \rangle
+ {\cal O} \left( g^4 \right)\, ,
\end{equation}
whose evaluation leads to
\begin{equation}
\label{MultiEoneGluon}
{\cal G} (\vec n_1, \dots , \vec n_j) = \frac{1}{2 \pi}
\frac{1}{\sin^{j + 2} \chi_1}
\prod_{i = 2}^{j}
\delta ( \cos \chi_1 - \cos \chi_i ) \delta (\varphi_1 - \varphi_i)
\int d \mbox{\boldmath$k$}{}^2
\left( \sqrt{\mbox{\boldmath$k$}{}^2} \right)^{j}
\, \rho_{\rm PT} \left( \mbox{\boldmath$k$}{}^2 \right) ,
\end{equation}
where $(\chi_i, \varphi_i)$ are the spherical coordinates of the unit
vector $\vec n_i$. Notice that this correlator vanishes unless
all vectors are aligned. Higher order corrections, discussed below,
will relax this condition.

Starting from Eq.\ (\ref{MultiEoneGluon}), we easily calculate the
total flow into the right or left hemispheres. According
to (\ref{MomentsEREL}), it is given by
\begin{equation}
\label{eLL}
\langle
{\cal E}_L^N
\rangle
=
\langle
{\cal E}_R^N
\rangle
=
\frac{2}{N} \int_0^{{\mit\Lambda}^2} d \mbox{\boldmath$k$}{}^2
\left( \sqrt{\mbox{\boldmath$k$}{}^2} \right)^{N}
\, \rho_{\rm PT} \left( \mbox{\boldmath$k$}{}^2 \right)
= 2 C_F \frac{\alpha_s}{\pi} \frac{{\mit\Lambda}^N}{N^2} \, ,
\end{equation}
where in the last relation we give the result for fixed
coupling. The vanishing of inter-hemisphere correlations,
\begin{equation}
\label{eLR}
\langle
{\cal E}_R^N {\cal E}_L^M
\rangle = 0 + {\cal O} \left( \alpha_s^2 \right) \, ,
\end{equation}
implies that the shape function, $f (\varepsilon_R,\varepsilon_L)$, is 
factorized into the product of two functions, each depending on a single 
energy variable
\begin{equation}
\label{Factor}
f_{\rm one-gluon} (\varepsilon_R, \varepsilon_L)
= f (\varepsilon_R) f (\varepsilon_L) \, ,
\end{equation}
with the function defined as $f (\varepsilon_i) = \langle \delta (
\varepsilon_i - {\cal E}_i ) \rangle$ for $i = R, L$.  Nonfactorizable 
corrections to Eq.\ (\ref{Factor}) begin with $\alpha_s^2$-contributions 
to multiple energy flow correlators ${\cal G} ( \vec n_1, \dots , \vec n_j )$ 
associated with final states with two real gluons. Phenomenological analyses
\cite{KorTaf00,GarRat01} have shown that power corrections to certain event 
shapes, in particular the heavy-jet mass, are very sensitive to these 
corrections.

{\bf 4. Order $\alpha_s^2$ and correlations between hemispheres.}
To study correlations between hemispheres at lowest order, $\alpha_s^2$, in
${\cal G}(\vec n_1, \vec n_2)$, we restrict ourselves to Feynman diagrams in 
Fig.\ \ref{GreenFigure} involving two-gluon cuts. According to the color
structure, one can separate abelian and nonabelian contributions, ${\cal G} =
{\cal G}_{A} + {\cal G}_{NA}$.

The abelian part, ${\cal G}_{A} (\vec n_1, \vec n_2)$, arises from QED-like
diagrams. Due to the exponentiation properties of eikonal lines
\cite{Ste81}, their contribution is factorized into the product of 
independent single-gluon emissions,
\begin{equation}
\label{Abelian}
{\cal G}_{A} (\vec n_1, \vec n_2)
= \frac{\alpha_s^2}{4 \pi^4} \frac{C_F^2}{\sin^3\! \chi_1 \sin^3\! \chi_2}
\int_0^{{\mit\Lambda}^2}
\frac{d \mbox{\boldmath$k$}{}_1^2}{\sqrt{\mbox{\boldmath$k$}{}_1^2}}
\frac{d \mbox{\boldmath$k$}{}_2^2}{\sqrt{\mbox{\boldmath$k$}{}_2^2}}
= {\cal G} (\vec n_1) {\cal G} (\vec n_2) \, ,
\label{abelian}
\end{equation}
where ${\cal G} (\vec n)$ is given by Eq.\ (\ref{1G}) and
where we impose a cutoff, ${\mit\Lambda}$,
on gluon transverse momenta, as above.  
The overall factor $(\sin\chi_1\, \sin\chi_2)^{-3}$
corresponds to two independent collinear enhancements,
when either gluon is emitted nearly parallel to one
of the eikonal directions.
For now, we study ${\cal G}_{A} (\vec n_1, \vec n_2)$ at fixed $\alpha_s$.
We will comment afterward on the role of the infrared behavior
of the running coupling.

The nonabelian part, ${\cal G}_{NA} (\vec n_1, \vec n_2)$, receives
contributions from the crossed ladder, self-energy and triple-gluon
vertex diagrams (see, e.g.,\ Fig.\ 1 in Ref.\ \cite{Bel98}). The total 
expression is a somewhat complicated function depending on the spherical 
angles $(\chi_1, \varphi_1)$ and $(\chi_2, \varphi_2)$. Closer 
examination reveals, however, that ${\cal G}_{NA} (\vec n_1, \vec n_2)$ 
depends, up to a prefactor, only on two specific combinations of these 
angles
\begin{equation}
{\cal G}_{NA} (\vec n_1, \vec n_2) =
\frac{\alpha_s^2}{\pi^2}
\frac{C_F C_A}{\sin^3 \chi_1 \sin^3 \chi_2}
{\mit\Lambda}^2
{\cal F} \bigg( \cosh (\eta_2 - \eta_1),
\cos (\varphi_2 - \varphi_1) \bigg) \, ,
\label{nonabelian}
\end{equation}
with
\begin{equation}
\cosh ( \eta_2 - \eta_1 )
= \frac{1}{2} \tan \frac{\chi_1}{2} \cot \frac{\chi_2}{2}
+ \frac{1}{2} \tan \frac{\chi_2}{2} \cot \frac{\chi_1}{2} \, ,
\end{equation}
where the rapidity $\eta_i$ is defined in Eq.\ (\ref{Rapidity}). This 
property is a consequence of boost invariance along the $z$-direction, 
and rotational invariance around it. The explicit expression for the 
dimensionless function ${\cal F}$ is rather cumbersome, and in fact we 
do not need it for our purposes. Indeed, the correlator ${\cal G}$ enters 
into the expression for the moments of the shape function 
(\ref{MomentsEREL}) under the integral over azimuthal angles $\varphi_i$. 
Since the weights $w_{R,L}$ (\ref{ERL}) do not depend on the $\varphi_i$, 
we may integrate over the azimuthal angles, to arrive at
\begin{equation}
\label{Nonabelian}
\int_{0}^{2 \pi} d \varphi_1 \int_{0}^{2 \pi} d \varphi_2 \
{\cal G}_{NA} ( \vec n_1, \vec n_2 ) = \frac{\alpha_s^2}{\pi^2}
\frac{C_F C_A}{\sin^3 \chi_1 \sin^3 \chi_2}
{\mit\Lambda}^2 \bar{ \cal F} (|\eta_2 - \eta_1|) \, ,
\end{equation}
where
\begin{equation}
\bar{ \cal F} (\eta)
= \frac{1}{\left( 2 {\mit\Lambda} \right)^2}
\int_0^{{\mit\Lambda}^2}
\frac{d \mbox{\boldmath$k$}{}_1^2}{\sqrt{\mbox{\boldmath$k$}{}_1^2}}
\frac{d \mbox{\boldmath$k$}{}_2^2}{\sqrt{\mbox{\boldmath$k$}{}_2^2}}
\frac{1
+ D \left( \mbox{\boldmath$k$}{}_1^2, \mbox{\boldmath$k$}{}_2^2 \right)}{
D^2 \!\left( \mbox{\boldmath$k$}{}_1^2, \mbox{\boldmath$k$}{}_2^2 \right)}
\left\{
D \!\left( \mbox{\boldmath$k$}{}_1^2, \mbox{\boldmath$k$}{}_2^2 \right)
\left( \coth \eta - 1 \right)
-
\frac{\sqrt{\mbox{\boldmath$k$}{}_1^2
\mbox{\boldmath$k$}{}_2^2}}{\mbox{\boldmath$k$}{}_1^2
+ \mbox{\boldmath$k$}{}_2^2 } \frac{1}{\sinh \eta}
\right\} \, ,
\end{equation}
with the function $D$ defined as
\begin{equation}
D \left( \mbox{\boldmath$k$}{}_1^2, \mbox{\boldmath$k$}{}_2^2
\right)
= 1 + 2 \frac{\sqrt{\mbox{\boldmath$k$}{}_1^2
\mbox{\boldmath$k$}{}_2^2}}{\mbox{\boldmath$k$}{}_1^2
+ \mbox{\boldmath$k$}{}_2^2 } \cosh \eta \, .
\end{equation}
Integrating over transverse momenta, we find
\begin{eqnarray}
\bar{ \cal F} (\eta) \!\!\!&=&\!\!\!
\left( \coth \eta - 1 \right)
\left( 2 + \eta \cosh \eta \coth \eta \right)
+
\frac{\cosh \eta }{2 \sinh^4 \eta}
\left(
2 \eta \cosh^2 \eta - 3 \eta
+ \sinh \eta
\right) \nonumber\\
&&\hspace{10mm} -
\left( \coth \eta - 1 + \frac{1}{\sinh \eta} \right)
\ln (2 + 2 \cosh \eta)\, .
\label{calFdef}
\end{eqnarray}
The function $\bar{ \cal F}$ depends only on the relative rapidity, $\eta = 
|\eta_1 - \eta_2|$, of the emitted gluons. We find that at large $\eta$ it 
decreases exponentially
\begin{equation}
\label{Flarge}
\bar{ \cal F} (\eta) = 2 \eta \, {\rm e}^{- 3 \eta}
+ {\cal O} \left( {\rm e}^{- 4 \eta} \right)\, ,
\end{equation}
while at small $\eta$ it has a pole,
\begin{equation}
\label{Fsmall}
\bar{ \cal F} (\eta) = 
\frac{1}{\eta} \left\{ \frac{49}{12} - 4 \ln 2 \right\} +
{\cal O} \left( \eta^0 \right).
\end{equation}
These properties may be understood as follows.

When $\chi_k \rightarrow 0, \pi$, the $k$-th gluon is emitted collinear 
to one of the eikonal directions.  At first sight, the nonabelian
contribution, Eq.\ (\ref{nonabelian}) has the same double collinear
enhancement as the abelian term, (\ref{abelian}) from 
the factor $(\sin\chi_1\, \sin\chi_2)^{-3}$.  The large-$\eta$
behavior $\exp\left( - 3 \eta \right)$ in $\bar{ \cal F}(\eta)$, however,
prevents enhancements when the directions of the gluons are widely
separated in rapidity.  The nonabelian term therefore has at most
a single collinear enhancement from configurations where both gluons 
approach the direction of one of the eikonal lines.
This result is consistent with the exponentiation
of collinear logarithms in inclusive cross sections, which
requires that the double-logarithmic collinear enhancement at two loops 
be given by the square of the enhancement at one loop.

At small rapidity intervals, $|\eta_2 - \eta_1| \to 0$, at fixed 
$\eta_k \ne 0, \pi$, the gluons propagate in the same direction. 
This gives rise to another collinear pole, $1/\eta$,
from (\ref{Fsmall}), which is compensated by a virtual ${\cal O}
(\alpha_s)$ correction to single gluon emission.  Once combined, they
define a distribution regularized via a  `plus'-prescription,
and give no new collinear enhancement to any infrared safe
cross section.

Let us examine the contribution of the correlator ${\cal G} (\vec n_1, 
\vec n_2)$ to the total energy flow into the left and/or right hemispheres,
Eqs.\ (\ref{eLL}) and (\ref{eLR}). We find that (\ref{eLL})
receives an ${\cal O} (\alpha_s^2)$ correction, while (\ref{eLR})
develops a non-zero value. The leading order $\alpha_s^2$-contribution to the
correlator $\langle {\cal E}_R^M {\cal E}_L^N \rangle$ is naturally
decomposed into an abelian ($\sim C_F^2$) and nonabelian ($\sim C_F C_A$) 
part. Due to the factorization property, (\ref{Abelian}), the former is 
reduced to the product of two factors, $\langle
{\cal E}_R^M \rangle \langle {\cal E}_L^N \rangle$, each given by
the lowest order expression (\ref{eLL}). The nonabelian piece
defines the irreducible part of the correlator $\langle {\cal E}_R {\cal E}_L
\rangle$,
\begin{equation}
\langle\langle {\cal E}_R {\cal E}_L \rangle\rangle
=
\langle {\cal E}_R {\cal E}_L \rangle
-
\langle {\cal E}_R \rangle \langle {\cal E}_L \rangle .
\end{equation}
It is given by the real-gluon contribution to ${\cal G}_{NA} (\vec n_1,
\vec n_2)$ alone, evaluated in Eq.\ (\ref{Nonabelian}). Because $\vec n_1$ 
and $\vec n_2$ belong to different hemispheres, the virtual gluon 
contribution 
is non-vanishing only for $\chi_1 = \chi_2 = \pi/2$.  This corresponds to a
single point in the phase space, $0 \leq \chi_1 \leq \pi/2$ and $\pi/2 \leq 
\chi_2 \leq \pi$, and vanishes upon integration over $\chi_{1,2}$.
Substituting the result of the perturbative computation of Eqs.\
(\ref{Nonabelian}) and (\ref{calFdef}), one finds after some algebra
\begin{equation}
\label{CorrelationLR}
\langle\langle
{\cal E}_R {\cal E}_L
\rangle\rangle
= \frac{\alpha_s^2}{\pi^2} C_F C_A {\mit\Lambda}^2
\int_0^\infty d \eta \, \eta \, {\rm e}^{- \eta} \bar{ \cal F} (\eta)
\approx \frac{\alpha_s^2}{\pi^2} C_F C_A \, {\mit\Lambda}^2 \cdot 0.5314 \, .
\end{equation}
This correlator describes cross-talk between the two hemispheres and it 
is positive. As we will see below, this has direct consequences for the 
differential event shape distributions, confirmed by phenomenological 
analysis \cite{KorTaf00}.

So far, we have kept the couplings fixed in our investigation of
$\langle\langle {\cal E}_R {\cal E}_L \rangle\rangle$.
The inclusion of higher order logarithmic corrections,
of course, requires a resummation, and we certainly do not
regard the numerical value of the correlation in  (\ref{CorrelationLR})
as a prediction. The positive sign of $\langle\langle {\cal E}_R 
{\cal E}_L \rangle\rangle$, however, is stable under any resummation 
that incorporates long-distance effects through the running of the 
couplings with transverse momenta, and even
as functions of the relative rapidity $\eta$. This follows from the 
form of integrand of Eq.\ (\ref{calFdef}), which is positive-definite 
for any fixed values of the transverse momenta and $\eta$. Similarly,
the explicit $\chi_i$-dependence in (\ref{nonabelian}) is a direct 
consequence of boost invariance, and holds both perturbatively and 
nonperturbatively.

{\bf 5.  From the correlators to the shape function.}
We are now ready to apply the expressions that we have obtained
for the energy flow correlators to reconstruct the shape function 
$f (\varepsilon_R, \varepsilon_L)$.
We start with the single-gluon approximation, in which
the factorization property (\ref{Factor}) holds, and 
for which the correlations of
energy flow are given by Eqs.\ (\ref{eLL}) and (\ref{eLR}). It follows
from (\ref{Factor}) that
\begin{equation}
\label{Fourierf}
f (\varepsilon_R)
= \langle \delta \left( \varepsilon_R - {\cal E}_R \right) \rangle
=
\int_{- \infty}^{\infty} \frac{d \lambda}{2 \pi} \, {\rm e}^{i \lambda
\varepsilon_R}
\,
\left\langle
\exp \left( - i \lambda {\cal E}_R \right)
\right\rangle\, ,
\end{equation}
and similarly for $f (\varepsilon_L)$. To calculate the expectation
value entering this expression, we use the cumulant expansion,
\begin{equation}
\label{CumulantExp}
\langle {\rm e}^X \rangle
= \exp \sum_{j = 1}^{\infty}
\frac{1}{j!} \langle\langle X^j \rangle\rangle\, ,
\end{equation}
where the lowest cumulants are,
$\langle\langle 1 \rangle\rangle \equiv \langle 1 \rangle$,
$\langle\langle 12 \rangle\rangle \equiv \langle 12 \rangle - \langle 1
\rangle \langle
2 \rangle$, $\langle\langle 123 \rangle\rangle \equiv \langle 123 \rangle -
\langle 1
\rangle \langle 23 \rangle - \langle 12 \rangle \langle 3 \rangle - \langle 13
\rangle \langle 2 \rangle + 2 \langle 1 \rangle \langle 2 \rangle \langle 3
\rangle$, and analogously for higher cumulants
 (see, for example \cite{Kam80}). In introducing the cumulant 
expansion, we note the close resemblance to the classic analysis of particle 
multiplicity, \cite{Mue71}.

The shape function $f (\varepsilon_R)$ is found by substituting $X = - i 
\lambda {\cal E}_R$ in (\ref{CumulantExp}). In the one-gluon approximation, 
$\langle\langle {\cal E}^N \rangle\rangle = \langle {\cal E}^N \rangle$. 
Then, using the correlators (\ref{eLL}), we easily resum the series in 
(\ref{Fourierf}) and obtain
\begin{equation}
\label{OneGluonFull}
f (\varepsilon_R) = \int_{- \infty}^{\infty} \frac{d \lambda}{2 \pi}
\exp \left\{
i \lambda \varepsilon_R + \int_0^{{\mit\Lambda}^2} d \mbox{\boldmath$k$}{}^2 \,
\rho_{\rm PT} \left( \mbox{\boldmath$k$}{}^2 \right)
\int_{0}^{\infty}
d \eta \ \left(
{\rm e}^{- i \lambda
\sqrt{\mbox{\boldmath${\scriptstyle k}$}{}^2} {\rm e}^{- \eta} }
- 1
\right)
\right\} ,
\end{equation}
and analogously for $f (\varepsilon_L)$. It is straightforward to
verify that $f (\varepsilon_{R,L})$ is a positive definite distribution. 
The function we have just calculated describes a resummed single-gluon
contribution to the shape function. Its lowest-order expansion in
powers of $\alpha_s$ reproduces a familiar
$1/\varepsilon_R$ energy distribution,
\begin{equation}
f (\varepsilon_R) = \frac{1}{\varepsilon_R}
\theta \left( {\mit\Lambda} - \varepsilon_R \right)
\int_{\varepsilon_R^2}^{{\mit\Lambda}^2}
d \mbox{\boldmath$k$}{}^2 \,
\rho_{\rm PT} \left( \mbox{\boldmath$k$}{}^2 \right)
+ {\cal O} \left( \alpha_s^2 \right) .
\end{equation}
At small $\varepsilon_R$, higher-order corrections become important,
and the asymptotic behavior of the resummed expression is drastically
changed. To see this, we notice that the $\varepsilon_R \to 0$ behavior
of (\ref{OneGluonFull}) is determined by the asymptotic behavior of the 
integrand as $\lambda \to \infty$. In this limit, using $\left.\int_0^\infty 
d \eta \, \left\{ \exp\left( - i \lambda \sqrt{\mbox{\boldmath$k$}{}^2} 
{\rm e}^{- \eta} \right) - 1 \right\} \right|_{\lambda \to \infty} 
\propto - \ln \lambda$, we find
\begin{equation}
\label{Small}
f (\varepsilon_R) \stackrel{\varepsilon_R \to 0}{\sim}
\varepsilon_R^{a_{\rm PT} - 1} ,
\qquad
a_{\rm PT} =
\int_0^{{\mit\Lambda}^2} d \mbox{\boldmath$k$}{}^2 \,
\rho_{\rm PT} \left( \mbox{\boldmath$k$}{}^2 \right) .
\end{equation}
Here $a_{\rm PT}$ has the interpretation  of the  number of particles 
per unit rapidity interval.  Perturbation theory thus suggests that $f$ 
vanishes as a power of the energy, although it
cannot reliably predict the absolute
value of the exponent $a_{\rm PT}$. 
In particular, $a_{\rm PT}$ suffers from infrared renormalon
ambiguities related to the divergence of the coupling constant at small
values of $\mbox{\boldmath$k$}{}^2$. One can estimate $a_{\rm NP} \approx
1 - 3$, replacing $\rho_{\rm PT}$ by a phenomenologically motivated
$\rho_{\rm NP}$ \cite{Web94}.

Now let us consider slightly larger values of $\varepsilon_R$,
for which Eq.\ (\ref{OneGluonFull}) is dominated by finite values
of $\lambda$, that is, values such that $\lambda \sqrt{\kappa^2}$
is a number of order unity for all $\kappa^2\le\Lambda^2$.
In this region, we may expand the exponential,  $\exp \left( - i 
\lambda \sqrt{\mbox{\boldmath$k$}{}^2} {\rm e}^{- \eta} \right)$
as a power series in  $\lambda$. If we keep only the linear term 
in $\lambda$, we get
\begin{equation}
f^{(0)} (\varepsilon_R) \propto \delta 
\left( \varepsilon_R - \langle {\cal E}_R
\rangle \right) \, ,
\end{equation}
where the superscript refers to the linear approximation used in
(\ref{OneGluonFull}).  This model for the shape function is
equivalent to a ``shift" \cite{IRR} of the perturbative 
distribution in Eq.\ (\ref{convolution}).

Clearly, a simple shift cannot readily be combined with the
low-$\varepsilon_R$ behavior of Eq.\ (\ref{Small}).  On
the other hand, the shift results from keeping only
the lowest order in $\lambda$ in our expansion of
the exponent in Eq.\ (\ref{OneGluonFull}).
The inclusion of the next order, $\lambda^2$, 
term amounts to a smearing of the delta-function into a Gaussian form
\begin{equation}
\label{Single}
f^{(1)} (\varepsilon_R) \approx \frac{1}{ \sqrt{ 2 \pi \langle\langle
{\cal E}_R^2 \rangle\rangle } } \exp \left\{
-
\frac{\bar\varepsilon_R^2}{2 \langle\langle {\cal E}_R^2 \rangle\rangle}
\right\} ,
\qquad\qquad
\bar \varepsilon_{R}
\equiv \varepsilon_{R} - \langle {\cal E}_{R} \rangle ,
\end{equation}
where $\langle\langle {\cal E}_R^2 \rangle\rangle = \langle{\cal E}_R^2
\rangle
+ {\cal O} (\alpha_s^2)$, with the first term given by (\ref{eLL}).

At order $\lambda^2$, we may also begin
to take into account the effects of the correlations between
hemispheres, Eq.\ (\ref{CorrelationLR}), in the shape function 
(\ref{ShapeFuncTwo}). Using the integral representation for the 
delta-function 
entering (\ref{ShapeFuncTwo}), and applying the expansion (\ref{CumulantExp}) 
we get, for $\varepsilon_{R,L} \sim {\mit\Lambda}$,
\begin{equation}
f (\varepsilon_R, \varepsilon_L)
\approx
\int_{- \infty}^\infty 
\frac{d\lambda_L}{2 \pi} \, \frac{d\lambda_R}{2 \pi} \;
{\rm e}^{i(\lambda_L\varepsilon_L+\lambda_R\varepsilon_R)}\;
\exp\, 
\left\{ 
- i \langle\langle 
\lambda_L {\cal E}_L + \lambda_R {\cal E}_R 
\rangle\rangle
- \frac{1}{2}
\langle\langle 
\left(
\lambda_L {\cal E}_L + \lambda_R {\cal E}_R
\right)^2
\rangle\rangle 
\right\} \, ,
\end{equation}
where we have neglected contributions of higher cumulants. In this region, 
we find the approximate behavior
\begin{eqnarray}
\label{Full}
f (\varepsilon_R, \varepsilon_L)
\approx
\frac{1}{2 \pi} \frac{1}{\sqrt{{\cal D}}}
\exp \left\{
- \frac{1}{2 {\cal D}}
\left(
\langle\langle {\cal E}_L^2 \rangle\rangle \bar\varepsilon_R^2
+
\langle\langle {\cal E}_R^2 \rangle\rangle \bar\varepsilon_L^2
-
2 \langle\langle {\cal E}_R {\cal E}_L \rangle\rangle
\bar\varepsilon_R \bar\varepsilon_L
\right)
\right\} \, .
\end{eqnarray}
Here the dispersion is defined as
\begin{equation}
\label{dispersion}
{\cal D} =
\langle\langle {\cal E}_L^2 \rangle\rangle
\langle\langle {\cal E}_R^2 \rangle\rangle
-
\langle\langle {\cal E}_R {\cal E}_L \rangle\rangle^2 ,
\end{equation}
and clearly $\langle\langle {\cal E}_R^2 \rangle\rangle = \langle\langle
{\cal E}_L^2 \rangle\rangle$. For small $\varepsilon_{R,L}$, the correlator
$\langle\langle {\cal E}_R {\cal E}_L \rangle\rangle$ does not affect the
asymptotic behavior of the shape function. Using (\ref{Small}) one finds $f
(\varepsilon_R, \varepsilon_L) \sim f (\varepsilon_R)f(\varepsilon_L)
\sim\left( \varepsilon_R\varepsilon_L \right)^{a_{\rm NP} - 1}$ as
$\varepsilon_{R,L}\to 0$. Together, these results imply, almost uniquely, 
the parameterization, Eq.\ (\ref{KTansatz}) for the double-hemisphere 
shape function $f (\varepsilon_R,\varepsilon_L)$, as a minimal form, 
neglecting higher cumulants.  The parameter $a$ of (\ref{KTansatz}) has 
the interpretation of the particle multiplicity per unit rapidity 
radiated by the boost-invariant sources, while $b$ measures the extent 
to which radiation in one hemisphere, produced perturbatively or otherwise, 
``spills over" into the adjoining hemisphere, increasing correlations 
between $\varepsilon_R$ and $\varepsilon_L$.

The Gaussian fall-off of (\ref{KTansatz}), of course, need not extend to 
arbitrarily large values of $\varepsilon_{L,R}$. For example,
the large-$\varepsilon_{R,L}$ behavior that follows from
the one-gluon approximation Eq.\ (\ref{OneGluonFull})
is actually $\sim \exp\left\{- \left(\varepsilon/{\mit\Lambda} 
\right) \, \ln \left(\varepsilon/{\mit\Lambda}\right) \right\}$,
which decreases somewhat less rapidly than the Gaussian in Eq.\
(\ref{KTansatz}), although faster than an exponential. From a 
phenomenological point of view, however, this difference should 
have a modest effect on convolutions like Eq.\ (\ref{convolution})
for physical cross sections.

To estimate the effect produced on the shape function (\ref{Full}) by 
nonzero correlations between the right and left hemispheres, we examine 
two extreme cases, $\langle\langle {\cal E}_R {\cal E}_L \rangle\rangle
= \langle\langle {\cal E}_L^2 \rangle\rangle$ and $\langle\langle 
{\cal E}_R {\cal E}_L \rangle\rangle=-\langle\langle {\cal E}_L^2 
\rangle\rangle$, corresponding to maximal and minimal value of the 
correlator, respectively. Since in both cases the dispersion 
(\ref{dispersion}) vanishes, the shape function is reduced to
$\delta(\varepsilon_R-\varepsilon_L)$ and
$\delta(\varepsilon_R+\varepsilon_L-2
\langle {\cal E}_{R} \rangle)$, correspondingly, indicating that the energy
flows into the two hemispheres are strongly correlated with each other. For
$\langle\langle {\cal E}_R {\cal E}_L \rangle\rangle > 0$ this correlation is
positive -- as $\varepsilon_R$ increases, the same is true for
$\varepsilon_L$. For $\langle\langle {\cal E}_R {\cal E}_L \rangle\rangle 
< 0$ the situation is opposite -- the energy flows into two hemispheres 
are anti-correlated with each other.

It follows from our calculation, Eq.~(\ref{CorrelationLR}), that the
correlations between two hemispheres are positive. This property is in 
agreement with phenomenological fits of a Gaussian-type ansatz to 
experimental data \cite{KorTaf00} on event shapes. It is also in accord with 
the dispersive approach \cite{DokMarWeb99}.

{\bf 6. Conclusions.} The event shape functions acquire an 
interpretation of energy flow distributions through hemispheres 
at infinities separated by the plane orthogonal to the jet 
axis. We have used perturbative QCD as a tool in our analysis of these 
functions, supplemented by conventional assumptions on the correspondence
between perturbative and nonperturbative behavior \cite{IRR}. 
While the resulting rapidity dependence is realistic, the
transverse momentum dependence is governed by nonperturbative physics. 
The resumation of multi-energy correlators, in a one-gluon approximation,
results in a power fall-off as $\varepsilon_{R,L} \to 0$. The sign of the 
correlation between the left and right hemispheres is more general than 
its perturbative approximation, and is a manifestation of the
tendency of radiation into one hemisphere to spread into the other.
As such, it is sensitive to the hadronization dynamics of soft
radiation at wide angles from the jets. The functional form of the shape 
functions resulting from our approach is in the agreement with recent
phenomenological analyses of differential event shape distributions
\cite{KorTaf00,GarRat01}.  We expect to extend our results to all orders 
in the underlying perturbative expansion.

\section*{Acknowledgements}

We are most grateful to Yu.\ Dokshitzer and A.\ Kaidalov for interesting
discussions. A.B.\ is grateful to D.\ Schiff for the hospitality extended to 
him at the Laboratoire de Physique Th\'eorique in Orsay. G.S.\ would like 
to thank Brookhaven National Laboratory for its hospitality. The work of 
A.B.\ and G.S.\ was supported in part by the National Science Foundation, 
grant PHY9722101. The work of G.K. was supported in part by the EU network
`Training and Mobility of Researchers', contract FMRX--CT98--0194.

\end{document}